\documentclass[a4paper,onecolumn,11pt,unpublished]{quantumarticle}
\pdfoutput=1
\usepackage[utf8]{inputenc}
\usepackage[english]{babel}
\usepackage[T1]{fontenc}
\usepackage{amsmath, braket, amsthm, amssymb, nicefrac, graphicx, subcaption}
\usepackage{hyperref}

\usepackage{tikz}
\usepackage{lipsum}

\usepackage{algorithm}
\usepackage{algpseudocode}
\algrenewcommand\algorithmicrequire{\textbf{Inputs:}}

\usepackage{algorithmicx}
\usepackage{algpseudocode}

\algdef{SE}[DOWHILE]{Do}{doWhile}{\algorithmicdo}[1]{\algorithmicwhile\ #1}%

\newtheorem{theorem}{Theorem}
\newtheorem{lemma}[theorem]{Lemma}
\newtheorem{proposition}[theorem]{Proposition}
\newtheorem{assumption}[theorem]{Assumption}

\title{Parsimonious Optimisation of Parameters in Variational Quantum Circuits}

\author{Sayantan Pramanik}
\affiliation{Robert Bosch Centre for Cyber-Physical Systems, Indian Institute of Science}
\affiliation{Corporate Incubation, TATA Consultancy Services, Bengaluru, India}
\email{sayantanp@iisc.ac.in}

\author{Chaitanya Murti}
\affiliation{Robert Bosch Centre for Cyber-Physical Systems, Indian Institute of Science}
\email{mchaitanya@iisc.ac.in}
\author{M Girish Chandra}
\affiliation{TCS Research, TATA Consultancy Services, Bengaluru, India}
\email{m.gchandra@tcs.com}

\begin{document}

\maketitle

\begin{abstract}
Variational quantum circuits characterise the state of a quantum system through the use of parameters that are optimised using classical optimisation procedures that typically rely on gradient information. The circuit-execution complexity of estimating the gradient of expectation values grows linearly with the number of parameters in the circuit, thereby rendering such methods prohibitively expensive. In this paper, we address this problem by proposing a novel Quantum-Gradient Sampling algorithm that requires the execution of at most two circuits per iteration to update the optimisable parameters, and with a reduced number of shots. Furthermore, our proposed method achieves similar asymptotic convergence rates to classical gradient descent, and empirically outperforms gradient descent, randomised coordinate descent, and SPSA. 
\end{abstract}

\section{Introduction}\label{sec:intro}
The quantum processors available today are limited in terms of the number of qubits available, their coherence times, noise-resilience, etc. This phase has been dubbed as the Noisy, Intermediate-Scale Quantum era \cite{nisq}. Variational Quantum Algorithms (VQAs) \cite{vqe, qaoa, ccq, transfer, ibm_spsa} - which employ both quantum and classical processors to work in tandem - investigate and utilise their potential for solving various problems in Chemistry, Machine Learning and Optimisation. In VQAs, we operate under the premise that a variational circuit represented by $U(\boldsymbol{\theta})$ acts on the conventional initial state $\ket{0}^{\otimes n}$ of the qubits, to obtain the state $\ket{\psi(\boldsymbol{\theta})}$ $= U(\boldsymbol{\theta})\ket{0}^{\otimes n}$; for the sake of brevity, the superscript $\otimes n$ will henceforth be dropped. We assume the circuits consist of single-qubit Pauli rotation gates, along with 2-qubit $CX$s, which together form a universal set of quantum logic gates. The rotational gates are parameterised by the list of parameters $\boldsymbol{\theta}\in\mathbb{R}^k$. The objective of variational algorithms is to find $\boldsymbol{\theta}^*$ such that:
\begin{equation}
    \boldsymbol{\theta}^* = \arg \min_{\boldsymbol{\theta}} \braket{\psi(\boldsymbol{\theta})|H|\psi(\boldsymbol{\theta})}
\end{equation}
for a Hermitian observable $H$. To simplify notation, we use:
\begin{equation}\label{eq:mu}
    \mu_{\boldsymbol{\theta}} = \braket{\psi(\boldsymbol{\theta})|H|\psi(\boldsymbol{\theta})}
\end{equation}

The current state-of-the-art method for estimating quantum gradients of expectation values for use in first-order methods for variational quantum algorithms is the parameter-shift rule (PSR) \cite{psr, mitarai}. PSR is equivalent to the finite-difference method, except that individual parameters are perturbed by finite values $s_i$, thus giving us a formula for gradients of single-qubit Pauli rotations as:
\begin{equation}
    \frac{\partial\mu_{\theta_i}}{\partial \theta_i} = \frac{\mu_{\theta_i+s_i}-\mu_{\theta_i-s_i}}{2}
\end{equation}
where $s_i=\nicefrac{\pi}{2}$. However, PSR requires running the quantum circuits (for a certain number of shots) twice, with $\theta_i + s_i$ and $\theta_i - s_i$, respectively to obtain the gradients with respect to each individual parameter. Thus, the total number of circuits evaluated is $2k$ at each iteration of the optimization routine, which prohibits scalability.

Several recent works have acknowledged the importance of reducing the number of quantum measurements, and have developed techniques for shot-frugal optimisation \cite{icans, rosalin, sgd, sglbo}. However, the number of measurements is calculated as the product of circuits executed and the shots they were run for. This is relevant because many cloud-providers that provision access to gate-model quantum processors employ separate pricing strategies for each circuit that is executed, and the number of shots it is executed for. Table \ref{table:price} provides an overview of the cost incurred in running each circuit and each shot on various quantum processors. Evidently, the per-circuit price appears to be 10-1000 times more expensive than per-shot price.

\begin{table}[h]
\centering
\begin{tabular}{|c|c|c|} 
 \hline
 \textbf{Processor} & \textbf{Per-circuit price (USD)} & \textbf{Per-shot price (USD)} \\ \hline \hline
 IonQ - Harmony & 0.3 & 0.01 \\ \hline
 IonQ - Aria & 0.3 & 0.03 \\ \hline
 OQC - Lucy & 0.3 & 0.00035 \\ \hline
 Rigetti - Aspen-M & 0.3 & 0.00035 \\ \hline
\end{tabular}
\caption{Per-circuit and per-shot prices for running circuits on real gate-based quantum processors through a popular cloud service provider.}
\label{table:price}
\end{table}

As such, we feel that the field of minimising the number of circuit-executions is underexplored, with only a few methods proposed, the most widely used and studied of which is the classical technique of Simultaneous Perturbation Stochastic Approximation (SPSA) \cite{ibm_spsa, icvqa, spsa1, spsa2}. SPSA also promises two circuit-evaluations for parameter-updation per iteration with perturbed parameters, and uses them to estimate the slope in the random direction \cite{icans}. However, it has been reported to be sensitive to the choice of hyperparameters \cite{icans}, and as shown later in Sec. \ref{sec:results}, not found to perform well for Quantum Machine Learning (QML) tasks. Besides gradient descent (based on PSR), and SPSA, the results from QGSA have also been compared against randomised coordinate descent \cite{rcd}, which randomly picks a parameter and updates it based on the gradient-estimate with respect to that parameter, resulting in two circuit runs. But it has been observed from the results in Sec. \ref{sec:results} that both SPSA and RCD end up consuming more iterations to converge.

To address these issues, we introduce the Quantum-Gradient Sampling Algorithm (QGSA) which requires at most two circuits to be executed (with a reduced number of shots) to update the parameters at each iteration. Furthermore, we show that it has the same asymptotic rate of convergence as gradient descent (GD). In practice, it was found to perform better than the aforesaid methods at a fraction of the number of measurements, as demonstrated in Sec. \ref{sec:results}.

\section{The Quantum Gradient Sampling Algorithm}\label{sec:methodology}

In this section, we state Algorithm \ref{alg:gs}, our proposed Quantum-Gradient Sampling Algorithm. The algorithm uses only two circuit evaluations to update the parameters in the variational quantum circuit, in contrast to the $2k$ evaluations necessitated by methods that estimate the gradients. It does this by utilising Theorem \ref{thm:bound} and sampling random vectors from a bounded probability distribution.\\\\

\noindent We begin by stating the following assumption.
\begin{assumption}\label{assump}
Going forward, without any loss of generality, we assume that the eigenvalues of the Hermitian observable $H$ lie in $[0,1]$.
\end{assumption}
\noindent We note that $H$ may be expressed as a linear combination of Pauli operators $P_i$ with real coefficients $c_i$:
\begin{equation}
    H = \sum_i c_iP_i
\end{equation}
The observable $H$ may be normalised as follows to obtain another Hermitian observable $H'$ whose eigenvalues lie in the range $[0,1]$\footnote{S.P. would like to thank Dr. Sourav Chatterjee and Dr. Anirban Mukherjee, from the Corporate Incubation team at Tata Consultancy services, for helpful discussions regarding Hamiltonian normalisation.}:
\begin{equation}
    H' = \frac{1}{2}\left( \frac{H}{\sum_i c_i} + I \right)
\end{equation}
\\\\
Our algorithm differs from classical gradient descent algorithms since instead of estimating the gradient, a random vector that is sufficiently close to the gradient is used as a descent direction. This yields an iterative algorithm, where the parameters ${\boldsymbol\theta}^{(t)}$ at the $t^{th}$ iteration are updated as:
\begin{equation}\label{eq:update1}
    {\boldsymbol\theta}^{(t+1)} = \arg \min (\mu_{{\boldsymbol\theta}^{(t)}-\alpha_t \boldsymbol{g_s}}, \mu_{{\boldsymbol\theta}^{(t)}+\alpha_t \boldsymbol{g_s}})
\end{equation}
where $\alpha_t>0$ is the step-size, $\boldsymbol{g_s} \in \mathbb{R}^k$ is a random vector whose components are i.i.d and sampled from a bounded probability distribution, as per Theorem \ref{thm:bound}, (for instance, \textit{Uniform}$(-2\sqrt{\mu_{\boldsymbol\theta}},2\sqrt{\mu_{\boldsymbol\theta}})$). QGSA relies on the fact that the bounds on the distribution decrease as the value of the objective function decreases at each iteration. The update rule~\ref{eq:update1} can be written as
\begin{equation}\label{eq:update2}
    {\boldsymbol\theta}^{(t+1)} = {\boldsymbol\theta}^{(t)} - s_t\alpha_t \boldsymbol{g_s}
\end{equation}
with $s_t=-\text{sgn} (\boldsymbol{g_t}^\top \boldsymbol{g_s})$, and $\boldsymbol{g_t}$ being the actual gradient of $\mu_{\boldsymbol\theta}^{(t)}$. The update step executes only two circuits (with a reduced number of shots, as explained later in Sec \ref{sec:shots}), compared to $2k$ circuit executions required by methods that estimate the gradient using the parameter-shift rule \cite{psr}.
\noindent We now state the Quantum-Gradient Sampling Algorithm formally in Algorithm~\ref{alg:gs}.\\
\begin{algorithm}[h]
\caption{Quantum-Gradient Sampling Algorithm}\label{alg:gs}
\begin{algorithmic}
\Require initial point $\boldsymbol\theta^{(0)}$, $n_\mu$, $n_g$, $\epsilon$
\State \textbf{start}
\State $\boldsymbol\theta \leftarrow \boldsymbol\theta^{(0)}$
\Do
\State evaluate $\mu_{\boldsymbol{\theta}}$ with $n_\mu$ shots
\State $\boldsymbol{g_s} \sim$ $\mathbb{P}\left(-2\sqrt{\mu_{\boldsymbol{\theta}}},2\sqrt{\mu_{\boldsymbol{\theta}}}\right)$ \Comment{where $\mathbb{P}$ is a probability density function bounded between $-2\sqrt{\mu_{\boldsymbol{\theta}}}$ and $2\sqrt{\mu_{\boldsymbol{\theta}}}$}
\State $\alpha \leftarrow \cfrac{\lvert \boldsymbol{g_{\boldsymbol{\theta}}}^\top \boldsymbol{g_s} \rvert}{L\| \boldsymbol{g_s} \|_2^2}$ \Comment{where $\boldsymbol g_{\boldsymbol{\theta}}$ is the gradient, and $L$ is the Lipschitz smoothness constant of $\mu_{\boldsymbol{\theta}}$}
\State $\boldsymbol{\theta}^- \leftarrow {\boldsymbol\theta}-\alpha \boldsymbol{g_s}$
\State $\boldsymbol{\theta}^+ \leftarrow {\boldsymbol\theta}+\alpha \boldsymbol{g_s}$
\State evaluate $\mu_{\boldsymbol{\theta}^-}$ and $\mu_{\boldsymbol{\theta}^+}$ with $n_g$ shots
\State $\boldsymbol{\theta} = \arg \min (\mu_{\boldsymbol{\theta}^-}, \mu_{\boldsymbol{\theta}^+})$
\doWhile{$(\|\boldsymbol g_{\boldsymbol{\theta}}\|_2 \geq \epsilon$)} 
\State \textbf{return} $\boldsymbol{\theta}$
\State \textbf{end}
\end{algorithmic}
\end{algorithm} 
\!%

\noindent We also provide another variant of the algorithm for use in the common setting wherein access to the gradient $\boldsymbol{g_{\boldsymbol{\theta}}}$ is unavailable. This algorithm is stated formally in Algorithm~\ref{alg:prac_gs} in Appendix \ref{ap:alg}.

\subsection{Perturbation bounds for quantum gradients}
In this section, motivated by the need to use randomly sampled surrogates for gradients, we derive bounds on the gradients of expectation values in the quantum setting.\\\\

\noindent Before introducing bounds on the gradients of expectation values, we present the following helpful result:
\begin{lemma}\label{lemma1}
If the $i^{th}$ parameter $\theta_i$ of an ansatz $U(\boldsymbol{\theta})$, $i \in [k]$, is perturbed by a small quantity $\delta_i$, such that the corresponding state changes from $\ket{\psi}_{\theta_i}$ to  $\ket{\Tilde{\psi}}_{\theta_i+\delta_i}$, then $\lim_{\delta_i\to 0} \|\ket{\psi}_{\theta_i} - \ket{\Tilde{\psi}}_{\theta_i+\delta_i}\| = |\delta_i|$, where $\| \cdot \|$ is the $l_2$ norm of a vector.
\end{lemma}

\begin{proof}
We first define $\ket{v}$ such that $\ket{\psi} = U_{i+1:k}(\theta_{i+1:k})U_i(\theta_i)\ket{v}$, $\ket{\epsilon} = \ket{\Tilde{\psi}-\psi}$, and note that $U_i(\theta_i) = e^{-i\theta_i G_i}$. Then,
\begin{equation}
\begin{aligned}
\lim_{\delta_i\to 0} \braket{\epsilon | \epsilon} & = \lim_{\delta_i\to 0} \left(\braket{\Tilde\psi | \Tilde\psi} +\braket{\psi | \psi}-\braket{\Tilde\psi | \psi}-\braket{\psi | \Tilde\psi} \right)\\
 & = \lim_{\delta_i\to 0} \left( 2-\bra{v}U_i^\dagger(\theta_i+\delta_i)U_i(\theta_i)\ket{v}-\bra{v}U_i^\dagger(\theta_i)U_i(\theta_i+\delta_i)\ket{v} \right) \\
 &= \lim_{\delta_i\to 0} \left( 2-\bra{v}(e^{-i\delta_i G_i}+e^{i\delta_i G_i})\ket{v} \right) \\
 &= \lim_{\delta_i\to 0} \left( 2-\bra{v}(2I\cos{\delta_i})\ket{v} \right) \;\;\; \mathrm{since \; for }\; G^2=I, \; e^{i\phi G} = I\cos \phi +iG\sin \phi\\
 &= \lim_{\delta_i\to 0} 4 \sin^2{\frac{\delta_i}{2}} = \delta_i^2 
\end{aligned}
\label{eq:lemma1}
\end{equation}
\end{proof}

That is, if one of the parameters of a variational quantum circuit is perturbed by $\delta_i$, then the magnitude of change in the corresponding state of the circuit is equal to the absolute value of the perturbation. In Theorem~\ref{thm:bound}, we use this observation to derive non-trivial bounds on the derivative of $\mu_{\boldsymbol{\theta}}$.

\begin{theorem}\label{thm:bound}
The partial derivative of $\mu_{\boldsymbol{\theta}}$ (as defined in Eq. \eqref{eq:mu}) with respect to $\theta_i$ lies in the range $[-2\sqrt{\mu_{\boldsymbol\theta}}, 2\sqrt{\mu_{\boldsymbol\theta}}]$, for all $i \in [k]$.
\end{theorem}

\begin{proof}
Continuing with the notation in lemma \ref{lemma1}, we have:
\begin{equation}
    \braket{\tilde\psi|H|\tilde\psi} = \braket{\psi|H|\psi}+\braket{\epsilon|H|\psi}+\braket{\psi|H|\epsilon}+\braket{\epsilon|H|\epsilon},
\end{equation}
where the last term is of the order of $\delta_i^2$, and may be ignored, resulting in 
\begin{equation}
    \mu_{\theta_i+\delta_i}-\mu_{\theta_i} = \braket{\epsilon|H|\psi}+\braket{\psi|H|\epsilon}
\end{equation}
Applying Cauchy-Schwarz inequality and noting that $\braket{\psi|H^2|\psi}\leq  \braket{\psi|H|\psi}$, we obtain:
\begin{equation}
-2\sqrt{\mu_{\boldsymbol\theta}}\|\epsilon\| \leq \mu_{\theta_i+\delta_i}-\mu_{\theta_i} \leq 2\sqrt{\mu_{\boldsymbol\theta}}\|\epsilon\|
\end{equation}
Finally, using the result of lemma \ref{lemma1}, it is straightforward to see that
\begin{equation}\label{eq:bounds}
    -2\sqrt{\mu_{\boldsymbol\theta}} \leq \frac{\partial\mu_{\theta_i}}{\partial \theta_i} \leq 2\sqrt{\mu_{\boldsymbol\theta}}
\end{equation}
\end{proof}

We note that the partial derivative of the expectation value with respect to each parameter is a function of the expectation value itself. The elements of the random vector are sampled from a probability distribution bounded as per Eq. \eqref{eq:bounds}. As a result, in the case of minimisation problems, if the function value decreases in consecutive iterations, then the bounds on the gradients get tighter, and the approximation of the gradient with the random vector gets better. We ensure this through Lemma \ref{lem:descent}, where we find bounds on the step-size that guarantees a decrease in objective function value from one iteration to the next.

\subsection{Quantum-Gradient Sampling Algorithm}\label{sec:gs}
In this section, we analyze the convergence of Algorithm~\ref{alg:gs}. In Theorem~\ref{thm:convergence}, we observe that under Assumption \ref{assump}, and with the correct choice of step-size $\alpha_t$ at each iteration, our algorithm achieves an asymptotic convergence rate of $O(\nicefrac{1}{\sqrt{T}})$, which is equivalent to that of classical gradient descent, while requiring up to $\sim 2k$ fewer circuit-executions per iteration. \\\\
\noindent We begin by providing a bound on the step-size $\alpha_t$ that guarantees reduction in the objective function value. 

\begin{lemma}\label{lem:descent}
    At iteration $t$, there exists a step-size $\alpha_t$ which guarantees a decrease in the objective function value from $\mu^{(t)}$ to $\mu^{(t+1)}$ (as defined in Eq. \eqref{eq:mu}), corresponding to ${\boldsymbol\theta}^{(t)}$ and ${\boldsymbol\theta}^{(t+1)}$ (which are related as per Eq. \eqref{eq:update1}), where $\boldsymbol{g_s}$ is a random vector $\in \mathbb{R}^k$.
\end{lemma}
\begin{proof}
Let $\mathbf{H}(\boldsymbol\zeta)$ denote the Hessian of $\mu_{\boldsymbol\zeta}$ (with $\boldsymbol\zeta$ being a convex combination of ${\boldsymbol\theta}^{(t)}$ and ${\boldsymbol\theta}^{(t+1)}$), and $L$ its Lipschitz smoothness constant. From the Taylor expansion of $\mu^{(t+1)}$:
\begin{equation}\label{eq:Taylor}
    \begin{aligned}
        \mu^{(t+1)}-\mu^{(t)} &= s_t \alpha_t \boldsymbol{g_t}^\top \boldsymbol{g_s} + \frac{\alpha_t^2 s_t^2}{2} \boldsymbol{g_s}^\top \mathbf{H}(\boldsymbol\zeta) \boldsymbol{g_s} \\
        & \leq -\alpha_t \lvert \boldsymbol{g_t}^\top \boldsymbol{g_s} \rvert + \frac{\alpha_t^2 L}{2} \| \boldsymbol{g_s} \|_2^2 \\
    \end{aligned}
\end{equation}
Thus, we obtain the following range of $\alpha_t$ that guarantees reduction in function value:
\begin{equation}\label{eq:step}
    0 \leq \alpha_t \leq \frac{2\lvert \boldsymbol{g_t}^\top \boldsymbol{g_s} \rvert}{L\| \boldsymbol{g_s} \|_2^2}
\end{equation}
\end{proof}

\begin{theorem}\label{thm:convergence}
    If the step-size $\alpha_t$ at each iteration satisfies Eq. \eqref{eq:step}, and the parameters $\boldsymbol{\theta}$ are updated as per Eq. \eqref{eq:update1}, then the Quantum-Gradient Sampling Algorithm converges in $T$ iterations to a stationary point of $\mu_{\boldsymbol{\theta}}$ with a rate given by $O\left( \frac{1}{\sqrt{T}}\right)$.
\end{theorem}
\begin{proof}
Choosing $\alpha_t = \frac{2\lvert \boldsymbol{g_t}^\top \boldsymbol{g_s} \rvert}{aL\| \boldsymbol{g_s} \|_2^2}$ with $a>1$, Eq. \eqref{eq:Taylor} simplifies to:
\begin{equation}
    \mu^{(t+1)}-\mu^{(t)} \leq -\frac{2}{aL} \left(1-\frac{1}{a} \right) \|\boldsymbol{g_t}\|_2^2 \cos^2(\phi_t)
\end{equation}
Taking expectation on both sides, and summing over $t$,
\begin{equation}
\begin{aligned}
    \sum_{t=0}^{T-1}\mathbb{E}\left[\mu^{(t+1)}-\mu^{(t)}\right] &\leq -\frac{2}{aL} \left(1-\frac{1}{a} \right) \sum_{t=0}^{T-1} \|\boldsymbol{g_t}\|_2^2 \mathbb{E}\left[\cos^2(\phi_t)\right] \\
    \implies \mu^{(0)}-\mathbb{E}[\mu^{(T-1)}] &\geq \frac{2}{aL} \left(1-\frac{1}{a} \right) T \epsilon_g^2 \mathbb{E}\left[\cos^2(\phi_t)\right] \\
    \implies \mu^{(0)}-\mu^{*} &\geq \frac{2}{aL} \left(1-\frac{1}{a} \right) T \epsilon_g^2 \mathbb{E}\left[\cos^2(\phi_t)\right]
\end{aligned}
\end{equation}
where $\phi_t$ is the angle between $\boldsymbol{g_s}$ and $\boldsymbol{g_t}$ at iteration $t$, and $\epsilon_g$ is a lower bound on the norm of the gradient of $\mu$.
\end{proof}
The choice of an appropriate step-size, as per Eq. \eqref{eq:step}, requires knowledge of the gradient at any given iteration. To circumvent this issue, we set the step-size to a sufficiently low value, and demonstrate empirically in Sec. \ref{sec:results} that this still leads to an advantage over gradient-based methods.

\subsubsection{Requisite Number of Shots}\label{sec:shots}
The gradient-sampling method expends its two circuit evaluations in discerning the descent-direction at each iteration. We posit that this requires $\mu_{\boldsymbol{\theta}^+}$ and $\mu_{\boldsymbol{\theta}^-}$ to be evaluated with a lower precision, thus opening up the potential for using a lower number of shots, where $\boldsymbol{\theta}^+ := {\boldsymbol\theta}^{(t)}+\alpha_t \boldsymbol{g_s}$ and $\boldsymbol{\theta}^- := {\boldsymbol\theta}^{(t)}-\alpha_t \boldsymbol{g_s}$.
\begin{proposition}
    Evaluating $\mu$ (defined in Eq. \eqref{eq:mu}) with a precision of $\epsilon_\mu$ and a confidence of $1-\delta$ requires $n_\mu \geq \frac{1}{2\epsilon_{\mu}^2}\ln{\left(\frac{2}{\delta}\right)}$ shots.
\end{proposition}
\begin{proof}
    This follows directly from Hoeffding's inequality.
\end{proof}

\begin{proposition}
    Finding if $\boldsymbol{\theta}^+$ (defined above) is a descent direction, i.e., $\mu_{\boldsymbol{\theta}^+}\leq \mu^{(t)}$, with a confidence of $1-\delta$ requires $n_g \geq \frac{1}{2(\mu^{(t)}-\mu_{\boldsymbol{\theta}^+})^2}\ln{\left(\frac{2}{\delta}\right)}$ shots.
\end{proposition}
\begin{proof}
Let $\hat\mu_{\boldsymbol{\theta}^+}$ be an unbiased estimator of $\mu_{\boldsymbol{\theta}^+}$. A simple geometrical analysis reveals that if $|\hat\mu_{\boldsymbol{\theta}^+} - \mu_{\boldsymbol{\theta}^+}| \geq |\mu^{(t)}-\mu_{\boldsymbol{\theta}^+}|$, then we fail to correctly obtain the direction of descent. Using Hoeffding's inequality again to bound the probability of failure gives the necessary number of shots to be $n_g \geq \frac{1}{2(\mu^{(t)}-\mu_{\boldsymbol{\theta}^+})^2}\ln{\left(\frac{2}{\delta}\right)}$.
\end{proof}
Thus, when the difference in objective value between subsequent iterations is larger than $\epsilon_\mu$, gradient-sampling requires a lower number of shots than gradient-based methods. However, recent studies on stochastic gradient descent, and its variations, \cite{icans, rosalin, sgd, sglbo} show that even using a very low number of shots (as few as a single shot) to estimate the gradient can result in adequate performance. Hence, it might be possible to incorporate the same shot-frugality into QGSA to obtain a reduction both in the number of circuit-executions and shots.

\subsubsection{Termination Criteria}
Though QGSA claims to use two circuit-evaluations per iteration, in practice, one iteration may suffice. One may choose to evaluate either $\mu_{\boldsymbol{\theta}^+}$ or $\mu_{\boldsymbol{\theta}^-}$, and forgo the other if the corresponding $\boldsymbol{\theta}^+$ or $\boldsymbol{\theta}^-$ is found to be a descent-direction, further reducing the number of circuits to be executed. In case neither of $\boldsymbol{\theta}^+$ or $\boldsymbol{\theta}^-$ offers a reduction in the objective function value, the step-size may be diminished as $\alpha_{(t+1)} \leftarrow \nicefrac{\alpha_t}{(1+\gamma)}$, where $\gamma$ is a decay parameter $\geq 0$. If no reduction is obtained over a configurable number of iterations, for very low values of $\alpha_t$, or for satisfactorily low bounds on the probability distribution, the algorithm may be terminated with the belief of having reached a (local) minimum.

\subsubsection{Other Details}
Finally, we hypothesize that QGSA may be more resilient to noise than the first-order methods that estimate the gradient, as the effects of noise may be absorbed within the stochasticity of the sampled parameters at each iteration. The proposed algorithm may also help navigate barren plateaus \cite{barren} due to its stochastic nature, as suggested in \cite{sglbo}. QGSA might also be combined with other first-order methods, natural gradient descent \cite{qng}, and heuristics such as operator-grouping \cite{og1, og2, og3, og4, og5} (by replacing their gradient-estimation step), to enhance its performance.

\section{Application to Binary Classification}\label{sec:bc}
Binary classification is one of the earliest use-cases to have been attempted using Quantum Machine Learning, and has been studied extensively \cite{ccq, transfer, kernel, zz}. The motivation behind mentioning it here is to set up the context to describe the experiments, and results thereof, presented later in Sec. \ref{sec:results}, and to introduce a new loss function to address binary classification. Without delving too much into the details, a binary classification model consists of a hypothesis function/model $h_{\boldsymbol\theta}$, which depends on trainable parameters $\boldsymbol{\theta}$. For each data point $\boldsymbol{x}_i$, the function returns a prediction, which is then compared against the available ground truth/label $y_i$ through the use of a loss function $\mathcal{L}$. The overall empirical risk is defined as:
\begin{equation}
    \hat R_h := \sum_{i=1}^n \mathcal{L}\left(h_{\boldsymbol\theta}(\boldsymbol{x}_i), y_i\right) =: \mathcal{L}_{\boldsymbol{\theta}}
\end{equation}
A classical optimiser then adjusts the parameters to minimise the empirical risk to obtain a model that can be used to make predictions for unseen data points. It is important to mention that the hypothesis and loss functions make implicit assumptions and place a prior on the class-conditional density of the the data, and the promise of quantum computing in QML is to provide families of variational circuits/ansaetze that are classically difficult-to-simulate.

The QGSA, recalling from above, minimises functions of the form $\braket{\psi(\boldsymbol{\theta})|H|\psi(\boldsymbol{\theta})}$, where the eigenvalues of $H \in [0,1]$. To provide a natural fit between this and the loss function, we define a new loss as:
\begin{equation}\label{eq:loss}
\begin{aligned}
    \mathcal{L}_{QH}(h_{\boldsymbol\theta}(\boldsymbol{x}), y) &:= \braket{\psi(\boldsymbol\theta, \boldsymbol{x})|(I-y\mathcal{O})|\psi(\boldsymbol\theta, \boldsymbol{x})} \\
 &\;= \frac{(1-yh_{\boldsymbol\theta}(\boldsymbol{x}))}{2}
\end{aligned}
\end{equation}
where $h_{\boldsymbol\theta}(\boldsymbol{x}) = \braket{\psi(\boldsymbol\theta, \boldsymbol{x})|\mathcal{O}|\psi(\boldsymbol\theta, \boldsymbol{x})}$; $\mathcal{O}$ is a user-defined, configurable observable whose eigenvalues lie between $-1$ and $1$;  and consequently, those of $(I-y\mathcal{O}) \in [0,1]$ which satisfies Assumption \ref{assump}. The loss function proposed in Eq. \eqref{eq:loss} is very similar to the popularly used Hinge loss in classical ML, due to which we dub this as the Quantum Hinge (QH) loss. It must also be noted that restriction on eigenvalues of $H$ is not a necessary condition for QGSA to work (as Theorem \ref{thm:convergence} is independent of this assumption), but is nevertheless a nice-to-have since it lets us appeal to Theorem \ref{thm:bound} and obtain consistently-reducing bounds on the probability distribution. The QGSA algorithm, as such, works with arbitrary loss functions as exemplified in the subsequent section.

\begin{figure*}
\centering
\begin{subfigure}[t]{0.49\textwidth}
\centering
\includegraphics[width=\textwidth]{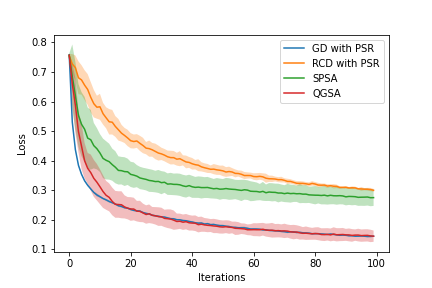}
\caption{Iris MSE}
\label{fig:iris_mse_1}
\end{subfigure}
\begin{subfigure}[t]{0.49\textwidth}
\centering
\includegraphics[width=\textwidth]{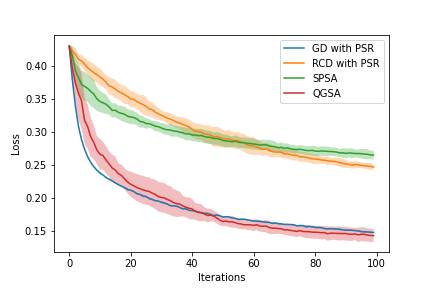}
\caption{Iris QH}
\label{fig:iris_qh_1}
\end{subfigure}
\begin{subfigure}[t]{0.49\textwidth}
\centering
\includegraphics[width=\textwidth]{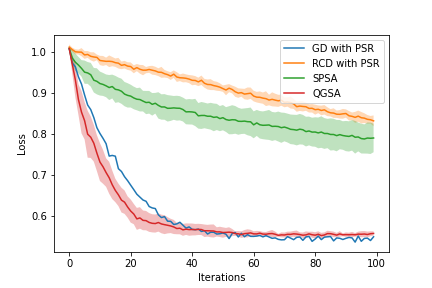}
\caption{Crack MSE}
\label{fig:crack_mse_1}
\end{subfigure}
\begin{subfigure}[t]{0.49\textwidth}
\centering
\includegraphics[width=\textwidth]{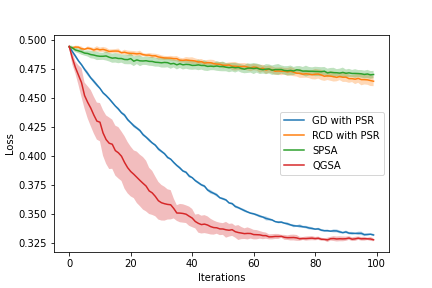}
\caption{Crack QH}
\label{fig:crack_qh_1}
\end{subfigure}
\begin{subfigure}[t]{0.49\textwidth}
\centering
\includegraphics[width=\textwidth]{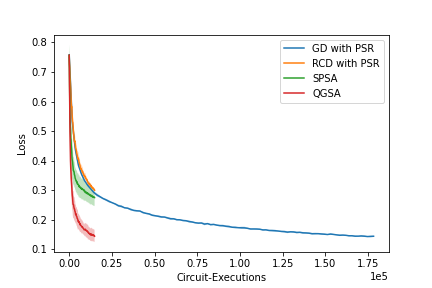}
\caption{Iris MSE}
\label{fig:iris_mse_2}
\end{subfigure}
\begin{subfigure}[t]{0.49\textwidth}
\centering
\includegraphics[width=\textwidth]{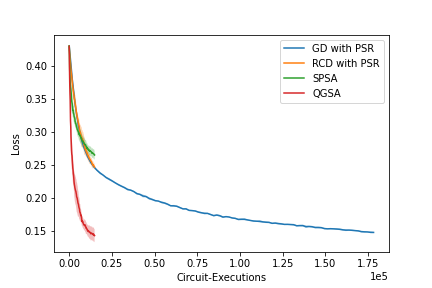}
\caption{Iris QH}
\label{fig:iris_qh_2}
\end{subfigure}
\begin{subfigure}[t]{0.49\textwidth}
\centering
\includegraphics[width=\textwidth]{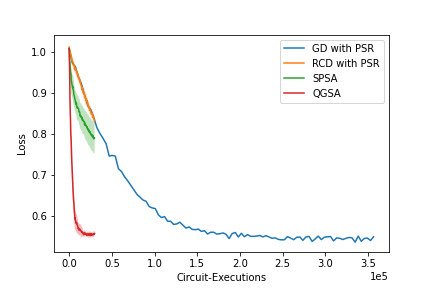}
\caption{Crack MSE}
\label{fig:crack_mse_2}
\end{subfigure}
\begin{subfigure}[t]{0.49\textwidth}
\centering
\includegraphics[width=\textwidth]{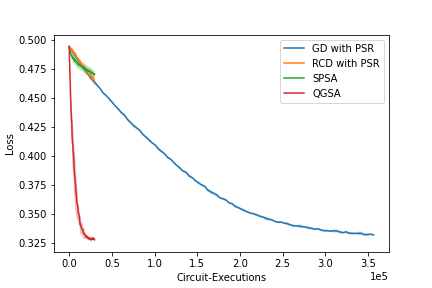}
\caption{Crack QH}
\label{fig:crack_qh_2}
\end{subfigure}
\caption{Plots of training losses on the y-axis against the number of iterations or the number of circuit-executions expended on updating the parameters on the x-axis, for both the Iris and Crack datasets. Sec. \ref{sec:expresults} provides details on the classifier used, and the caption of each individual subfigure identifies the data set and the loss function.}
\label{fig:results}
\end{figure*}

\section{Experimental Details and Results}\label{sec:expresults}
In this section, we provide details of the binary classification experiments that were conducted to investigate and compare the performance of QGSA against Gradient Descent (GD), Randomised Coordinate Descent (RCD), and SPSA, the former two of which use the PSR to estimate gradients, and the latter two use two function evaluations per iteration to update the parameters. The onus of this paper is not to propose good encoding schemes, ansaetze or loss functions for the problem being considered, or even how the trained (ML) models generalise to unseen data points, but to compare the rate of decrease in objective function value with respect to the number of circuits executed.

Binary classification was performed on the following two datasets:
\begin{itemize}
\item Iris dataset \cite{iris}: only the first two classes were considered and the features are translated and scaled to the range of $[0, \pi]$.
\item Kaggle Surface Crack Detection dataset \cite{crack}: The entire dataset consists of $20,000$ $227 \times 227$ RGB examples each of images with and without cracks. $100$ samples were randomly chosen from each class and preprocessed, as described in \cite{bmw}.
\end{itemize}

\noindent In both cases, the data points were labelled as $+1$ and $-1$ to denote the two classes.

\noindent The classification circuits were created by first encoding the data into the qubits using the $H$ gate, followed by the $R_Z$ gate with the features passed as parameters to the latter. The encoding layer was followed up with $3$ layers of the Basic Entangling Layers ansatz from Pennylane \cite{pennylane}, numbering the optimisable parameters in the circuit at $12$. Finally, $\braket{Z}$ of the first qubit in the circuit was measured, which played the role of $h_{\boldsymbol{\theta}}(\boldsymbol{x})$. The models with both QH and Mean Squared Error (MSE) as the loss functions, using the four aforementioned methods. In each case, the models were initiated with the same starting point and trained for $100$ iterations with step-size $\alpha=0.1$ (in case of SPSA, the values of the hyperparameters were: $a=0.1$, $\alpha=0.602$, $c=0.2$, and $\gamma=0.101$, which were chosen as prescribed in \cite{spsa2}, and $a$ was chosen to make the comparison fair against QGSA, GD, and RCD) on the noise-free simulator available through Pennylane \cite{pennylane}. In case of gradient-sampling, both $\mu_{\boldsymbol{\theta}^+}$ and $\mu_{\boldsymbol{\theta}^-}$ were evaluated, and keeping the stochasticity of the processes in mind, we report the results over $10$ trials.

\subsection{Results and Observations}\label{sec:results}

Fig. \ref{fig:results} depicts the plots of average training losses (over 10 experiments, along with their standard deviation) for binary classification of both the Iris and Crack datasets, using the classifiers detailed above. The training losses have been reported against the number of iterations of the optimisation procedure, as well as the total number of circuit executions used for updating the parameters. Both SPSA and RCD were found to require the same number of circuit evaluations on an average as GD, while QGSA consistently provided a dramatic reduction. To put this in perspective, training the classifier on the Crack dataset with QH loss using GD on a superconducting qubit based processor available through the aforementioned cloud service provider would consume $\sim$USD 230,000. In comparison, using QGSA, the same activity would require only about USD 18,000. Further, it may be observed that QGSA uses approximately the same number of iterations as GD, both of which perform a lot better than RCD and SPSA.

\section{Conclusion and Future Directions}
In this paper, we presented Quantum-Gradient Sampling Algorithm which uses random vectors drawn from bounded probability distributions, instead of estimating gradients, to update the parameters in variational quantum circuits. We also proved that QGSA has the same asymptotic rate of convergence as gradient descent, and demonstrated it's capability through binary classification on two different datasets. The results showed a $k$-fold reduction in the number of circuit executions on an average through the use of QGSA. The development of QGSA may be carried forward by incorporating it into first-order methods as a substitute for the gradient estimation step, and further reducing the requisite number of shots through shot-frugal methods. The performance of QGSA may also be investigated in navigating barren plateaus and providing resilience against noise.

\bibliography{qsga}
\bibliographystyle{unsrt}





  

\appendix
\section{Practical QGSA}\label{ap:alg}
\begin{algorithm}[h]
\caption{Practically implementable Quantum-Gradient Sampling Algorithm}\label{alg:prac_gs}
\begin{algorithmic}
\Require Step-size $\alpha$, step-size decay parameter $\gamma$, initial point $\boldsymbol\theta^{(0)}$, $n_\mu$, $n_g$, number of iterations $T$, $t=0$
\State \textbf{start}
\State $\boldsymbol\theta \leftarrow \boldsymbol\theta^{(0)}$
\For{$t\in[T]$}
\State evaluate $\mu^{(t)}$ with $n_\mu$ shots
\State $\boldsymbol{g_s} \sim$ $\mathbb{P}\left(-2\sqrt{\mu^{(t)}},2\sqrt{\mu^{(t)}}\right)$ \Comment{where $\mathbb{P}$ is a probability density function bounded between $-2\sqrt{\mu^{(t)}}$ and $2\sqrt{\mu^{(t)}}$}
\State $\boldsymbol{\theta}^- \leftarrow {\boldsymbol\theta}-\alpha \boldsymbol{g_s}$
\State evaluate $\mu_{\boldsymbol{\theta}^-}$ with $n_g$ shots
\If{$\mu_{\boldsymbol{\theta}^-} < \mu^{(t)}$}
    \State $\boldsymbol\theta \leftarrow \boldsymbol\theta^{-}$
\Else
    \State $\boldsymbol{\theta}^+ \leftarrow {\boldsymbol\theta}+\alpha \boldsymbol{g_s}$
    \State evaluate $\mu_{\boldsymbol{\theta}^+}$ with $n_g$ shots
    \If{$\mu_{\boldsymbol{\theta}^+} < \mu^{(t)}$}
    \State $\boldsymbol\theta \leftarrow \boldsymbol\theta^{+}$
    \Else
        \State $\alpha \leftarrow \frac{\alpha}{1+\gamma}$
    \EndIf
\EndIf
\EndFor
\State \textbf{return} $\boldsymbol{\theta}$
\State \textbf{end}
\end{algorithmic}
\end{algorithm}

\end{document}